\documentclass[12pt]{iopart}

\usepackage{amssymb}
\usepackage{xcolor,graphicx}

\begin{document}

\title{Hamiltonian closures for fluid models with four moments by dimensional analysis}

\author{M. Perin$^a$, C. Chandre$^a$, P.J. Morrison$^b$,  E. Tassi$^a$}
\address{$^a$ Aix-Marseille Universit\'e, Universit\'e de Toulon, CNRS, CPT UMR 7332, 13288 Marseille, France \\
$^b$ Department of Physics and  Institute for Fusion Studies, The University of Texas at Austin, Austin, TX 78712-1060, USA}
\ead{maxime.perin@cpt.univ-mrs.fr}

\begin{abstract}
Fluid reductions of the Vlasov-Amp\`ere equations that preserve the Hamiltonian structure of the parent kinetic model are investigated.  Hamiltonian closures using the first four moments of the Vlasov distribution are obtained, and all closures  provided by a dimensional analysis procedure for satisfying the Jacobi identity are identified. Two   Hamiltonian models emerge, for which the explicit closures are given, along with their Poisson brackets  and  Casimir invariants.  
\end{abstract}

\maketitle

\section{Introduction}
\label{Intro}

The Vlasov-Amp\`ere set of equations is a suitable framework for describing the dynamics of systems interacting through electrostatic forces.  In this work, we focus on the study of electrostatic plasmas even though the results may be applied to more general systems described in part by the Vlasov equation. We consider a one-dimensional plasma made of electrons of unit  mass and negative unit electric charge,  evolving in a neutralizing background of static ions. The evolution of the distribution function of the electrons $f$, defined on phase space with coordinates $(x,v)$,  and electric field $E$ is given by the Vlasov-Amp\`ere equations,  
\begin{eqnarray}
\label{eqVlasov}
\partial_t f&=-v\partial_x f+\widetilde{E}\partial_v f,\\
\label{eqPoisson}
\partial_t E&=-4\pi\widetilde{\jmath},
\end{eqnarray}
where $\widetilde{E}$ and $\widetilde{\jmath}$ are the fluctuating parts of the electric field $E$ and the current density $j=-\int vf\ \mathrm{d}v$ respectively.  We assume vanishing boundary conditions at infinity in the velocity $v$ so that integrals such as the charge and current densities are well-defined. In this work, we limit ourselves to the study of systems of unit length in the spatial domain $x$ with periodic boundary conditions.  The fluctuating part of the electric field is defined by 
$\widetilde{E}=E-\int_0^1 E\ \mathrm{d}x$.  The system is fully nonlinear, but has a form that builds in the preservation of the spatial average of $E$ and maintains momentum conservation.

The use of fluid reductions to describe the dynamics of a plasma is ubiquitous in plasma physics. Indeed, this usually allows one to decrease the complexity of the problem at hand and to gain physical insight into  the phenomenon under investigation since the dimension of phase space is reduced. Fluid reductions of the Vlasov-Amp\`ere equations are done by introducing fluid quantities such as the fluid moments
\begin{equation}
\label{defP}
P_n=\int v^nf(x,v,t)\ \mathrm{d}v.
\end{equation}
The associated dynamical equations are then obtained by multiplying Eq.~(\ref{eqVlasov}) by $v^n$ and integrating with respect to the velocity. This leads to
\begin{eqnarray}
\label{eqVlasovP}
\partial_t P_n&=-\partial_x P_{n+1}-nP_{n-1}\widetilde{E},\\
\label{eqPoissonP}
\partial_t E&=4\pi\widetilde{P_1},
\end{eqnarray}
for all $n\in\mathbb{N}$. In order for this system to be reduced, one has to truncate the infinite sequence  of Eq.~(\ref{eqVlasovP}). Truncating this system at order $N$, that is considering $(P_0,P_1,\dots,P_N,E)$ as dynamical field variables, one can see from Eq.~(\ref{eqVlasovP}) that the time evolution of $P_N$ depends on $P_{N+1}$. As a consequence,  it is necessary to  express $P_{N+1}$ in terms of  $(P_0,P_1,\dots,P_N,E)$ in order to close Eqs.~(\ref{eqVlasovP}) and (\ref{eqPoissonP}) and thereby obtain a fluid reduction.

Many models have been proposed based on as many closures with  various requirements (see, e.g., Refs.~\cite{Shadwick04,Shadwick05,Goswami05,Shadwick12}). A usual procedure consists in assuming a particular  form for the distribution function $f$ (e.g., Dirac, Maxwellian,\dots) depending on a finite number of parameters, and expressing the closure with respect to these parameters \cite{Braginskii65}.  Alternatively,   closures have been constructed  in order to recover certain kinetic effects \cite{Ott69,HKM,Hammett90,Hammett93,Sugama03,Passot04,Sarazin09}. In any event, a reduction by closure should be such that, if the parent model possesses a Hamiltonian structure \cite{Morrison80,Morrison82,Morrison98,Marsden02}, then the resulting fluid model should also have one, after discarding all the terms that  are supposed to provide dissipation. A closure procedure ignoring this aspect could potentially lead to the introduction of  some nonphysical dissipation~\cite{deGuillebon12,tronci}.  Consequently, here we  use  a procedure that preserves the Hamiltonian structure of the parent kinetic (Vlasov-Amp\`ere) system, which is one of its most important structural features. Specifically, in this work we present a model for the first four fluid moments of the distribution function, namely the density $\rho$, the fluid velocity $u$, the pressure $P$ and the heat flux $q$. This allows us to account for the time evolution of the heat flux, which is of great importance for the study of transport phenomena inside the plasma. For such a model with four moments, one has to find a closure for the fifth order moment of the distribution function, namely $P_4$. Here, we determine all the closures, obtained from a procedure based on dimensional analysis,  that preserve the Hamiltonian structure of the parent model~\cite{Morrison80b,Morrison82} given by Eqs.~(\ref{eqVlasov}) and (\ref{eqPoisson}). We show that there are only two such Hamiltonian closures.  The equations of motion of one of these two models are identical to the ones obtained with a bi-delta reduction \cite{Jin03,Gosse03,Chalons12}, i.e., assuming that the Vlasov distribution has the form
\begin{equation*}
f(x,v,t)=\omega_1 \delta (v-\mu_1)+\omega_2 \delta(v-\mu_2),
\end{equation*}
where $\omega_{1,2}$ and $\mu_{1,2}$ depend on space and time. 
It should be noted here that we obtain these equations without any assumption on the special form of the distribution function. We provide the explicit expressions of the Hamiltonian and the Poisson bracket for the two Hamiltonian models. In addition, we derive the global Casimir invariants, which are specific invariants resulting from the knowledge of the Poisson bracket. These conserved quantities can be used, e.g., to ensure the validity of a numerical simulation of the equations of motion.

The paper is organized as follows. In Sec.~\ref{Method} we describe the methodology used for the derivation of the two Hamiltonian reduced models. We start from the definitions of the appropriate variables, namely, the reduced fluid moments. Subsequently, we introduce our method, based on dimensional analysis, which leads to models  that obey the Jacobi identity. We show that there are only two such models.  In Sec.~\ref{secTwoModels}, we analyze the two resulting Hamiltonian closures, providing  explicit expressions for their Hamiltonians,   Poisson brackets,  and  Casimir invariants.

\section{Method}
\label{Method}

\subsection{Reduced moments}

Our purpose is to build a Hamiltonian fluid model for the first four moments of the distribution function, namely the density $\rho$, the fluid velocity $u$, the pressure $P$, the heat flux $q$ and the electric field $E$. These models will be referred to as 4+1 field   models,  where the 4 refers to the four first moments of the Vlasov distribution (or equivalently to $\rho$, $u$, $P$ and $q$) and the 1 refers to the electric field $E$. We begin by considering the Poisson structure of the parent model with $(f,E)$ as dynamical field variables. It was shown  in Ref.~\cite{Chandre13} that the system of Eqs.~(\ref{eqVlasov})-(\ref{eqPoisson}) possesses a Hamiltonian structure with Poisson bracket 
\begin{equation}
\label{brackVP}
\hspace{-.75 cm} \{F,G\}=\int f\left[\partial_x F_f\partial_v G_f-\partial_x G_f\partial_v F_f+4\pi(\widetilde{F_E}\partial_v G_f-\widetilde{G_E}\partial_v F_f)\right]\ \mathrm{d}x\mathrm{d}v,
\end{equation}
where $F_f$ (resp.\ $F_E$) denotes the functional derivative of $F$ with respect to $f$ (resp.\  $E$). In addition, Bracket~(\ref{brackVP}) is bilinear and satisfies the Leibniz rule and the Jacobi identity. The Hamiltonian of the system is given by
\begin{equation}
\label{hamVP}
\mathcal{H}=\int f\frac{v^2}{2}\ \mathrm{d}x\mathrm{d}v+\int\frac{E^2}{8\pi}\ \mathrm{d}x,
\end{equation}
where the first term accounts for the kinetic energy of the particles and the second one corresponds to the energy of the electric field.  Together with Bracket~(\ref{brackVP}), this Hamiltonian leads to Eqs.~(\ref{eqVlasov}) and (\ref{eqPoisson}) by using $\partial_t f=\{f,\mathcal{H}\}$ and $\partial_t E=\{E,\mathcal{H}\}$. We recall that such a bracket has Casimir invariants, i.e., functionals $C$ that  Poisson-commute with any other functionals of the Poisson algebra, $\{C,F\}=0$ for all $F$. Bracket~(\ref{brackVP}) has the following global (i.e., independent of the coordinates $x$ and $v$) Casimir invariants
\begin{eqnarray*}
&& C_1 = \int \varphi(f)\ {\rm d}x {\rm d}v,\\
&& C_2 = \int E\ {\rm d} x,
\end{eqnarray*}
for any scalar function $\varphi$, and a local Casimir invariant 
$$
C_{\rm L}=\partial_x E +4\pi\int f {\rm d}v,
$$
which is equivalent to Gauss's law.

The change from the kinetic to the fluid description is done by performing the change of variables defined by Eq.~(\ref{defP}) in Bracket~(\ref{brackVP}) and Hamiltonian~(\ref{hamVP}). 
The latter becomes
\begin{equation*}
\mathcal{H}=\frac{1}{2}\int\left(P_2+\frac{E^2}{4\pi}\right)\ \mathrm{d}x.
\end{equation*}
Making use of the chain rule to transform the functional derivatives, Bracket~(\ref{brackVP}) becomes \cite{Kup78,Gibbons81,Gibbons08}
\begin{equation}
\label{brackFluid}
\hspace{-.75 cm}\{F,G\}=\int j\left[P_{i+j-1}(G_j\partial_x F_i-F_j\partial_x G_i)+4\pi P_{j-1}(G_j\widetilde{F_E}-F_j\widetilde{G_E})\right]\ {\rm d}x,
\end{equation}
where $F_n$ denotes the functional derivative of $F$ with respect to $P_n$, and summation is implicit over the repeated indices $i$ and $j$.  Because  we want to construct a Hamiltonian model for the first four moments of the distribution function, we consider functionals of the kind $F[P_0,P_1,P_2,P_3,E]$. However,  the Poisson bracket~(\ref{brackFluid}) of two functionals of this kind depends explicitly on two additional moments, namely $P_4$ and $P_5$. In order to close the system, these two additional moments need to be expressed in terms of $P_{n\leq 3}$ and $E$. As a result, the Jacobi identity is no longer satisfied in general, and the resulting truncated and closed bracket is not of Poisson type. Consequently, the resulting system is not Hamiltonian, or in other terms, the reduction procedure potentially includes dissipation. We notice that the closure has to be performed on two moments, $P_4$ and $P_5$, which slightly differs from what has been stated in the introduction, concerning the closure performed on the equations of motion directly, where only one additional moment, $P_4$, needs to be closed. However we shall see in Sec.~\ref{hamCstr} that the expression of $P_5$ is entirely determined by   $P_4$.

We introduce the reduced fluid moments, which we find  to be more suitable variables for our purpose, 
\begin{equation}
\label{reducedMoments}
\rho=\int f\ \mathrm{d}v,\qquad u=\frac{1}{\rho}\int vf\ \mathrm{d}v,\qquad S_n=\frac{1}{\rho^{n+1}}\int(v-u)^nf\ \mathrm{d}v,
\end{equation}
for all $n\geq2$. The first and second ones correspond respectively to the usual density and fluid velocity. The higher-order moments are the central fluid moments with a specific scaling with respect to the density. The change from the usual fluid moments $P_n$ to the reduced fluid moments $(\rho,u,S_n)$, used hereafter,  is invertible so that the results, even though they are expressed in a different set of coordinates, are equivalent. This change is given by
\begin{equation*}
\rho=P_0,\qquad u=\frac{P_1}{P_0}, \qquad S_n=\frac{1}{P_0^{n+1}}\sum\limits_{m=0}^n {{n}\choose{m}}\left(\frac{-P_1}{P_0}\right)^{n-m}P_m,
\end{equation*}
for all $n\geq2$. The inverse of this transformation is given by
\begin{equation*}
P_0=\rho,\qquad P_1=\rho u, \qquad  P_n=\rho\left[u^n+\sum\limits_{m=2}^n {{n}\choose{m}}\rho^mu^{n-m}S_m\right].
\end{equation*}
Explicitly for the first four moments of the distribution function, this change of variables is given by
\begin{eqnarray*}
&\rho=P_0,  \hspace{ 2cm} &u=\frac{P_1}{P_0}, \\
&S_2=\frac{1}{P_0^3}\left(P_2-\frac{P_1^2}{P_0}\right),\qquad &S_3=\frac{1}{P_0^4}\left(P_3-3\frac{P_1P_2}{P_0}+2\frac{P_1^3}{P_0^2}\right),
\end{eqnarray*}
with the  inverse
\begin{equation*}
\hspace{-.5 cm}P_0=\rho,\quad P_1=\rho u,\quad P_2=\rho\left(u^2+\rho^2S_2\right),\quad P_3=\rho\left(u^3+3\rho^2 u S_2+\rho^3S_3\right).
\end{equation*}
In terms of the moments, Hamiltonian~(\ref{hamVP}) is  
\begin{equation}
\label{hamS}
\mathcal{H}=\frac{1}{2}\int\left(\rho u^2+\rho^3S_2+\frac{E^2}{4\pi}\right)\ \mathrm{d}x.
\end{equation}
The first part of Hamiltonian~(\ref{hamS}) accounts for the kinetic energy of the system while its second part corresponds to the internal energy. The last term, which  accounts for  the electric energy, remains unchanged compared to Eq.~(\ref{hamVP}). By considering functionals of the kind $F[\rho,u,S_2,S_3,E]$ and using the chain rule for the functional derivatives (see  \ref{appendixConstraints} for more details), Bracket~(\ref{brackVP}) takes the form
\begin{eqnarray}
\label{brackFourFields}
\{F,G\}&=&\int\Bigg[G_u\partial_x F_\rho-F_u\partial_x G_\rho+4\pi(G_u\widetilde{F_E}-F_u\widetilde{G_E})\nonumber \\
&-& \frac{1}{\rho}(G_u F_i-F_u G_i)\partial_x S_i +\alpha_{ij}\frac{F_i}{\rho}\frac{G_j}{\rho}+\partial_x\left(\frac{F_i}{\rho}\right)\beta_{ij}\frac{G_j}{\rho}\Bigg]\ \mathrm{d}x,
\end{eqnarray}
where $F_i$ denotes the functional derivative of $F$ with respect to $S_i$. From now on and unless otherwise stated, summation from 2 to 3 over repeated indices is implicit. The matrices $\alpha$ and $\beta$ have indices ranging  from 2 to 3 such that
\begin{equation}
\label{matAlphaBeta}
\hspace{-1.85 cm} \alpha=\partial_x\left(
\begin{array}{cc}
2S_3 & 2S_4-3S_2^2\\
3S_4-6S_2^2 & 3S_5 - 12S_2S_3
\end{array}
\right),
\quad
\beta=\left(
\begin{array}{cc}
4S_3 & 5S_4-9S_2^2\\
5S_4-9S_2^2 & 6S_5-24S_2S_3
\end{array}
\right).
\end{equation} 
We notice that $\partial_x\beta=\alpha+\alpha^t$, a property that ensures that Bracket~(\ref{brackFourFields}) is antisymmetric. From Definitions~(\ref{matAlphaBeta}) we see that the closure  requires reexpression of  $S_4$ and $S_5$, i.e., one has to express these two reduced moments with respect to the dynamical variables $(\rho,u,S_2,S_3,E)$ such that Bracket~(\ref{brackFourFields}) satisfies the Jacobi identity.

We remark that Bracket~(\ref{brackFourFields}) has several  subalgebras. Trivial ones include $F[\rho]$ (i.e., the algebra of functionals of the type $F[\rho]$), $F[u]$, $F[E]$, $F[\rho,E]$, and non-trivial ones include $F[\rho,u]$, $F[\rho,S_2,S_3]$, $F[u,E]$, $F[\rho,u,S_2,S_3]$, $F[\rho,u,E]$ and $F[\rho,S_2,S_3,E]$. The most interesting one is the subalgebra of functionals $F[\rho,S_2,S_3]$ for which $\rho$ becomes a Casimir invariant. The existence of this subalgebra is the reason for considering the reduced fluid moments $S_n$.  

\subsection{The Hamiltonian constraints}
\label{hamCstr}

In order to be a Poisson bracket, Bracket~(\ref{brackFourFields}) must  satisfy the Jacobi identity,  
\begin{equation*}
\{F,\{G,H\}\}+\{H,\{F,G\}\}+\{G,\{H,F\}\}=0.
\end{equation*}
Here we determine the conditions on $S_4$ and $S_5$ resulting from requiring the Jacobi identity. We begin by  assuming  that  $S_4$ and $S_5$ depend on $\rho$, $u$, $S_2$, $S_3$, $E$ and their derivatives $\partial_x^n \rho$, $\partial_x^n u$, $\partial_x^n S_2$, $\partial_x^n S_3$, $\partial_x^n E$ for $n$ lower than some order $\nu$. Using the  result obtained  in \ref{appendixJacobi}, we conclude that $S_4$ and $S_5$ do not depend on $\rho$, $u$, $E$ and their derivatives $\partial_x^n \rho$, $\partial_x^n u$, $\partial_x^n E$.  In addition, we show in \ref{appendixDerivatives} that in order for the Jacobi identity to be satisfied, we need to impose
\begin{equation*}
\gamma_{ljm}\gamma_{kij}=\gamma_{kjm}\gamma_{lij},
\end{equation*}
for all $i$, $k$, $l$ and $m$ ranging  from 2 to 3, where the summation is implicit on $j$, and
\begin{equation*}
\gamma_{ljm}(S_k,\partial_x S_k,\dots,\partial_x^{\nu-1} S_k)=\frac{\partial\alpha_{lj}}{\partial\partial_x^\nu S_m}.
\end{equation*}
For instance, for $l=2$, $m=3$, $i=2$ and $k=3$, we end up with $\gamma_{233}\gamma_{323}=0$ since $\gamma_{222}=0$ and $\gamma_{223}=0$ for $\nu\geq 2$. From Eq.~(\ref{matAlphaBeta}), we have $3\gamma_{233}=2\gamma_{323}$, therefore $\gamma_{233}=0$, or equivalently
\begin{equation*}
\frac{\partial S_4}{\partial\partial_x^{\nu-1} S_3}=0.
\end{equation*}
Using Eq.~(\ref{rnd55}) leads to
\begin{equation*}
S_3\frac{\partial\alpha_{23}}{\partial\partial_x^\nu S_2}=0.
\end{equation*}
Since this has to be true for any value of $S_3$, we thus conclude that $\gamma_{232}=0$, i.e., 
\begin{equation*}
\frac{\partial S_4}{\partial\partial_x^{\nu-1} S_2}=0.
\end{equation*}
Concerning $S_5$, Eq.~(\ref{rnd55}) for $l=i=3$ leads to
\begin{equation}
\label{eqKerBeta}
\beta_{kj}\frac{\partial S_5}{\partial\partial_x^{\nu-1} S_j}=0.
\end{equation}
There are two solutions to Eq.~(\ref{eqKerBeta}). The first solution is given by
\begin{equation*}
\frac{\partial S_5}{\partial\partial_x^{\nu-1} S_2}=0,\quad\frac{\partial S_5}{\partial\partial_x^{\nu-1} S_3}=0. 
\end{equation*}
The second solution requires $\det \beta~=0$, which, using Eq.~(\ref{matAlphaBeta}), can be written as 
\begin{equation*}
S_5=4S_2S_3+\frac{(5S_4-9S_2^2)^2}{4S_3}.
\end{equation*}
Since $S_4$ does not depend on $\partial_x^{\nu-1} S_2$ and $\partial_x^{\nu-1} S_3$, we again have
\begin{equation*}
\frac{\partial S_5}{\partial\partial_x^{\nu-1} S_2}=0,\quad\frac{\partial S_5}{\partial\partial_x^{\nu-1} S_3}=0. 
\end{equation*}
In what follows we will see that the second solution does not lead to a Hamiltonian closure. By induction on $\nu$ down to $\nu=2$ we show that $S_4$ and $S_5$ have to be functions of $S_2$ and $S_3$ only. These conditions are necessary but not sufficient, i.e., for any functions $S_4$ and $S_5$ of $S_2$ and $S_3$, Bracket~(\ref{brackFourFields}) does not satisfy the Jacobi identity in general. 

We compute in  \ref{appendixConstraints} the necessary and sufficient conditions on the closures for a fluid bracket of the type~(\ref{brackFourFields}) to satisfy the Jacobi identity. For four fluid moments, Eqs.~(\ref{constraintsJacobi1}) and (\ref{constraintsJacobi2}) are 
\begin{eqnarray}
\label{eqDS5}
\frac{\partial S_5}{\partial S_2}&=4S_3+\frac{\partial S_4}{\partial S_3}\left(\frac{\partial S_4}{\partial S_2}-3S_2\right),\\
\label{eqDS5bis}
\frac{\partial S_5}{\partial S_3}&=\left(\frac{\partial S_4}{\partial S_3}\right)^2+\frac{\partial S_4}{\partial S_2},\\
\label{eqS5}
6S_5&=4S_3\left(3S_2+\frac{\partial S_4}{\partial S_2}\right)-\frac{\partial S_4}{\partial S_3}\left(9S_2^2-5S_4\right).
\end{eqnarray}
We see from Eq.~(\ref{eqS5}) that the expression of $S_5$ is fully determined by $S_4$. By introducing the expression for $S_5$ given by Eq.~(\ref{eqS5}) into Eqs.~(\ref{eqDS5}) and (\ref{eqDS5bis}), we end up with the  following two  nonlinear second order partial differential equations:
\begin{eqnarray}
\label{PDEs1}
&4S_3\frac{\partial^2S_4}{\partial S_2^2}-\frac{\partial^2S_4}{\partial S_2\partial S_3}\left(9S_2^2-5S_4\right)-\frac{\partial S_4}{\partial S_2}\frac{\partial S_4}{\partial S_3}=12S_3,\\
\label{PDEs2}
&4S_3\frac{\partial^2S_4}{\partial S_3\partial S_2}-\frac{\partial^2S_4}{\partial S_3^2}\left(9S_2^2-5S_4\right)+12S_2=\left(\frac{\partial S_4}{\partial S_3}\right)^2+2\frac{\partial S_4}{\partial S_2}.
\end{eqnarray}
Provided that these two equations are satisfied, Bracket~(\ref{brackFourFields}) is a Poisson bracket and the resulting system is Hamiltonian. Solving these equations in general is  challenging;  consequently, in  what follows we restrict ourselves to the set of solutions provided by  dimensional analysis \cite{Gibbings11}. 

\subsection{Closures based on dimensional analysis}
\label{secDimAnalysis}

We consider all the closures for the fifth-order moment $S_4=g(S_2,S_3)$ that satisfy the constraints given by Eqs.~(\ref{PDEs1}) and (\ref{PDEs2}) based on a dimensional analysis argument. In order to proceed, we assume that the closure $S_4=g(S_2,S_3)$ does not depend on any further dimensional parameters. This would not be the case for, e.g., diffusion-like closures (Fourier's law, Fick's law, etc...) that  introduce phenomenological parameters resulting from various  hypotheses based on  characteristic scales of the dynamics of the system.  Indeed, in diffusion processes, diffusion coefficients  replace  information on the particle interactions, thus removing  small scale dynamics. Instead, we would like our reduction procedure to be very general and not to depend on the geometry of the system. Consequently, we seek  Hamiltonian closures where $S_4=g(S_2,S_3)$ do not depend on any further dimensional parameters. 

It can be shown from Eq.~(\ref{reducedMoments}) that the dimensions of the $S_n$'s, denoted $[S_n]$, are not independent. Indeed, for all $n\geq2$ we have $[S_n]=\mathrm{A}^n$,  where $\mathrm{A}=\mathrm{L}^2\mathrm{T}^{-1}$ with  $\mathrm{L}$ and  $\mathrm{T}$ denoting  the units of length and time, respectively.  As a consequence, the closure $S_4=g(S_2,S_3)$ involves three quantities and a unique physical dimension $\mathrm{A}$. Making use of the Buckingham $\pi$ theorem \cite{Gibbings11}, there exists two dimensionless quantities, denoted $\zeta$ and $\xi$, such that $S_4=g(S_2,S_3)$ reduces to $\zeta=R(\xi)$. Therefore,  this procedure eliminates one  of the  variables in the closure. Defining $\zeta=S_4/S_2^2$ and $\xi=S_3/S_2^{3/2}$ and inserting these expressions into Eqs.~(\ref{PDEs1})-(\ref{PDEs2}), we get the following two  constraints:
\begin{eqnarray}
\label{ODEs1}
&3\xi R''(6\xi^2+9-5R)+R'(3\xi R'-18\xi^2+R-9)+16\xi R=24\xi,\\
\label{ODEs2}
&R''(6\xi^2+9-5R)+R'(R'-5\xi)+4R=12.
\end{eqnarray}
To solve Eqs.~(\ref{ODEs1})-(\ref{ODEs2}), we compute their values for $\xi=0$. Defining  $R_0=R(0)$, $R'_0=R'(0)$, and $R''_0=R''(0)$,  Eqs.~(\ref{ODEs1}) and (\ref{ODEs2}) become
\begin{eqnarray}
\label{ODEs3}
&R'_0(R_0-9)=0,\\
\label{ODEs4}
&R''_0(9-5R_0)+R'^2_0+4R_0=12.
\end{eqnarray}
Equation~(\ref{ODEs3}) has two solutions: $R'_0=0$ and $R_0=9$. Equation~(\ref{ODEs4}) then reads $R''_0=4(3-R_0)/(9-5R_0)$ for $R'_0=0$ and $R'^2_0=12(3R''_0-2)$ for $R_0=9$. We now differentiate Eqs.~(\ref{ODEs1}) and (\ref{ODEs2}) with respect to $\xi$ and evaluate them at $\xi=0$. This gives us
\begin{eqnarray}
\label{ODEs5}
&R''_0(9-7R_0)+2R'^2_0+8R_0=12,\\
\label{ODEs6}
&R'''_0(9-5R_0)-R'_0(3R''_0+1)=0.
\end{eqnarray}
Using $R_0=9$, Eq.~(\ref{ODEs5}) together with Eq.~(\ref{ODEs4}) leads to $R''_0=-2/3$ and $R'^2_0=-48$. As a consequence, this solution is not real and of no interest for our purpose. The other solutions satisfy $(R_0,R'_0,R''_0,R'''_0)=(0,0,4/3,0)$ and $(R_0,R'_0,R''_0,R'''_0)=(1,0,2,0)$, where $R'''_0=R'''(0)$. Since the solution is unique for a given set of initial conditions, there exist only two solutions to Eqs.~(\ref{ODEs1}) and (\ref{ODEs2}). Moreover, one can see that $R(-\xi)$ is also a solution of these equations.   Consequently, the two solutions $R$  are even with respect to $\xi$. These two solutions are described in the next section.

\section{Hamiltonian fluid models with 4+1 fields}
\label{secTwoModels}

\subsection{Model without normal variables}

\begin{figure}
\centering
\includegraphics[width=0.5\textwidth]{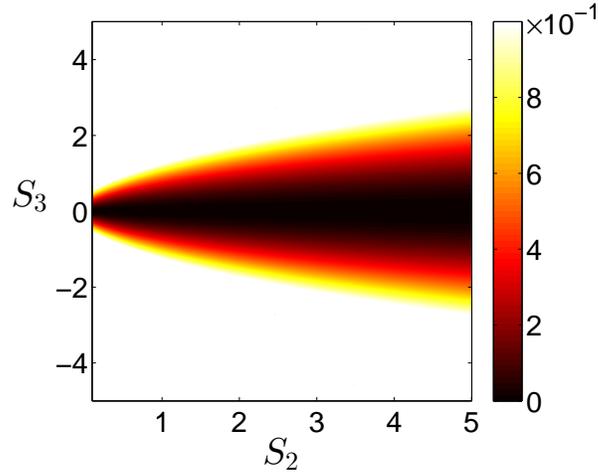}
\caption{Color map of $S_4$ (in $\mathrm{A}^4$ units as defined in Sec.~\ref{secDimAnalysis}) given by Eq.~(\ref{eqS4NonTrivial}) as a function of $S_2$ (in $\mathrm{A}^2$ units) and $S_3$ (in $\mathrm{A}^3$ units).}
\label{figS4NonTrivial}
\end{figure}

\begin{figure}
\centering
\includegraphics[width=0.5\textwidth]{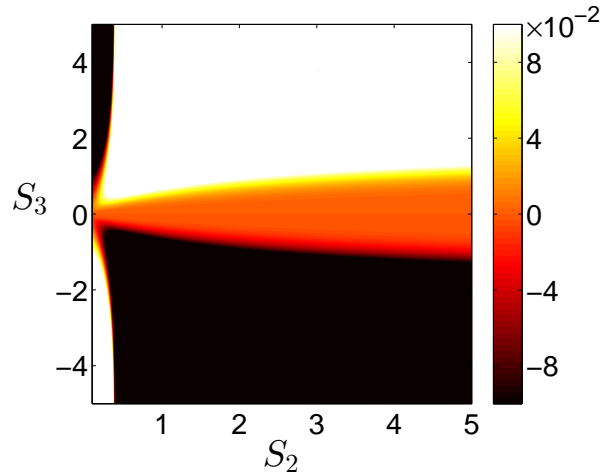}
\caption{Color map of $S_5$ (in $\mathrm{A}^5$ units as defined in Sec.~\ref{secDimAnalysis}) given by Eq.~(\ref{eqS5NonTrivial}) as a function of $S_2$ (in $\mathrm{A}^2$ units) and $S_3$ (in $\mathrm{A}^3$ units).}
\label{figS5NonTrivial}
\end{figure}

We consider the first solution to Eqs.~(\ref{ODEs1})-(\ref{ODEs2}),  corresponding to $(R_0,R'_0,R''_0,R'''_0)=(0,0,4/3,0)$. As mentioned in Sec.~\ref{secDimAnalysis}, the solution $R$ is even. Thus we introduce $R(\xi)=\bar{R}(\eta)$,  where $\eta=\xi^2$. Then, Eqs.~(\ref{ODEs1})-(\ref{ODEs2}) become
\begin{eqnarray}
\label{eq:Ab1}
&\xi\left[3\eta \bar{R}''(6\eta+9-5\bar{R})+\bar{R}'(9+3\eta \bar{R}'-7\bar{R})+4\bar{R}-6\right]=0,\\
\label{eq:Ab2}
&2\eta \bar{R}''(6\eta+9-5\bar{R})+\bar{R}'(9+2\eta \bar{R}'-5\bar{R}+\eta)+2\bar{R}-6=0.
\end{eqnarray}
By linearly combining these equations to eliminate terms in $\bar{R}''$, and introducing the new variable $\mu=-(\bar{R}-3\eta-9)/5$, we end up with an Abel equation of the second kind (see, for instance, Ref.~\cite{Polyanin03}):
\begin{equation*}
\mu\mu'-\mu=-\frac{6\eta+24}{25},
\end{equation*}
which has the  parametric solution 
\begin{equation}
\label{parametric}
\eta(\tau)=K\frac{(2-5\tau)^2}{(3-5\tau)^3}-4,\qquad 
\mu(\tau)=K\tau\frac{(2-5\tau)^2}{(3-5\tau)^3},
\end{equation}
where $K$ is some constant to be determined. Inserting Eq.~(\ref{parametric}) into Eqs.~(\ref{eq:Ab1})-(\ref{eq:Ab2}) implies these  equations are satisfied if and only if $K=27$. This leads to an explicit expression for the closure $\zeta=S_4/S_2^2=R(\xi)$ given by
\begin{equation}
\label{eq:R}
\hspace{-1 cm}R(\xi)=3\frac{[4+ 4 t(\xi)^2 - t(\xi) (\xi^2-8) - 8\xi^2][2 + 2 t(\xi)^2 - t(\xi) (\xi^2-2) - 4 \xi^2]}{[1-2\xi^2+3t(\xi)+t(\xi)^2]^2},
\end{equation}
where
$$
t(\xi)=\left(\frac{\sqrt{\xi^2(4+\xi^2)^3}-2-10\xi^2+\xi^4}{2}\right)^{1/3}.
$$
Furthermore, by using Eq.~(\ref{eqS5}), $S_5$ is given by $S_5=S_2S_3T(\xi)$ with 
\begin{equation}
\label{eq:T}
T(\xi)=2\frac{3\xi^2-R(\xi)^2-7R(\xi)}{R(\xi)-3\xi^2-9}.
\end{equation}
In summary, the Hamiltonian and the Poisson bracket resulting from this closure are given respectively by Eq.~(\ref{hamS}) and Eq.~(\ref{brackFourFields}) with $\alpha$ and $\beta$ given by Eq.~(\ref{matAlphaBeta}) with
\begin{eqnarray}
\label{eqS4NonTrivial}
& S_4=S_2^2R\left(\frac{S_3}{S_2^{3/2}}\right),\\
\label{eqS5NonTrivial}
& S_5=S_2S_3T\left(\frac{S_3}{S_2^{3/2}}\right),
\end{eqnarray}
where $R$ and $T$ are given by Eqs.~(\ref{eq:R})-(\ref{eq:T}). The dependence of the functions $R$ and $T$ in their arguments is not trivial. In order to help the reader visualize the closure relations corresponding to Eqs.~(\ref{eqS4NonTrivial})-(\ref{eqS5NonTrivial}), we provide, in Fig.~\ref{figS4NonTrivial} (resp. Fig.~\ref{figS5NonTrivial}), color maps showing the dependence of $S_4$ (resp. $S_5$) on $S_2$ and $S_3$.
As a side note, we remark that as $S_3$ goes toward 0, $S_4$ also goes to 0, as shown in Fig.~\ref{figS4NonTrivial}. Thus, with this closure, non-asymmetric distribution functions cannot exist. Consequently, the physical relevance of this solution is questionable.

In order to further characterize this Poisson bracket, we investigate its Casimir invariants. These are functionals $C[\rho,u,S_2,S_3,E]$ that  commute with any other functionals $F$, i.e., $\{F,C\}=0$ for all $F$. In particular,  $C$ commutes with the field variables. As a consequence, we must  impose
$$
\{\rho,C\}=-\partial_x C_u=0,
$$
which leads to
$$
C[\rho,u,S_2,S_3,E]=K_1\int u\ \mathrm{d}x+D[\rho,S_2,S_3,E],
$$
where $K_1$ is constant. Imposing that $C$ commutes with $u$ leads to
$$
\{u,C\}=-\partial_x D_\rho-4\pi\widetilde{D_E}+\frac{1}{\rho} D_i \partial_x S_i=0,
$$
whose solution is given by
$$
D[\rho,S_2,S_3,E]=K_2\int\rho \phi(S_2,S_3)\ \mathrm{d}x+K_3\int E\ \mathrm{d}x,
$$
where $K_2$ and $K_3$ are constant. Imposing that $C$ commutes with $S_i$ leads to
\begin{equation}
\label{eqK}
\{S_i,C\}=-\frac{K_1}{\rho}\partial_x S_i+K_2\frac{\alpha_{ij}}{\rho} \phi_j-\frac{K_2}{\rho}\partial_x\left(\beta_{ij} \phi_j\right)=0,
\end{equation}
where $\phi_i=\partial \phi/\partial S_i$. 
We then solve the associated homogeneous equation ($K_1=0$).  Again making use of the Buckingham $\pi$ theorem, we assume that there exist a real number $a$ and a function $\psi$ such that $\phi=S_2^a\psi(S_3/S_2^{3/2})$. The resulting equations are
\begin{eqnarray*}
&& 8 (a-1) a \xi \psi + \psi' \left( 3 - 18 a + 30 \xi^2 - 24 a \xi^2 - 3 R
+ 10 a R -      9 \xi R'\right)\\
&& \qquad\qquad\qquad +  3 \xi \left( 9 + 6 \xi^2 - 5 R\right) \psi'' = 0,\\
&& 2 a \psi + \psi' (-9 \xi + 4 a \xi + 3 R') + (-9 - 6 \xi^2 +      5 R) \psi'' = 0,\\
&& 4 a \psi [3 - 9 a + (-1 + 5 a)R -  3 \xi R'] =  3 \xi [\psi' (1 - 4 a + (-17 + 20 a) R\\
&& \qquad\qquad \qquad - 8 (a-1) T -       6 \xi R' + 6 \xi T') -    3 \xi (7 + 5 R - 4 T) \psi''],\\
&& 4 a \psi R' + \psi' [3 - 18 a + 5 (-3 + 2 a) R + 6 T - 6 \xi R' + 6 \xi T']\\
&& \qquad\qquad \qquad =  3 \xi (7 + 5 R - 4 T) \psi''.
\end{eqnarray*}
Combining the first two equations leads to
\begin{equation*}
2 a (4 a-1) \xi \psi  + [3 (1 - 6 a + \xi^2 - 4 a \xi^2) + (-3 + 10 a) R] \psi' = 0,
\end{equation*}
whose solution is given by
\begin{equation*}
\psi(\xi)=\psi_0 \exp\left(\int\limits^\xi\frac{2ay(4a-1)}{3(6a-1) +3y^2(4a-1) + R(y)(3 - 10 a)}\ \mathrm{d}y\right),
\end{equation*}
where $\psi_0$ is a constant. Inserting  this expression into  the previous equations provides the necessary constraints $a=0$ and $\psi(\xi)=\psi_0$. As a consequence, this model does not have Casimir invariants of the entropy-type \cite{Perin14}, i.e., of the form $\int \rho \phi(S_2,S_3)\ {\rm d} x$. Moreover, we can show in a similar way that Eq.~(\ref{eqK}) has no solution for the nonhomogeneous case ($K_1\neq0$), i.e.,   $K_1=0$ is required. The Poisson bracket resulting from this closure has only two global Casimir invariants, given by 
\[
 C_1 = \int\rho\ \mathrm{d}x \qquad\mathrm{and}\qquad 
 C_2 = \int E\ \mathrm{d}x,
\]
which are also Casimir invariants of the Vlasov-Amp\`ere equations. 

\subsection{Model with normal variables}

\begin{figure}
\centering
\includegraphics[width=0.5\textwidth]{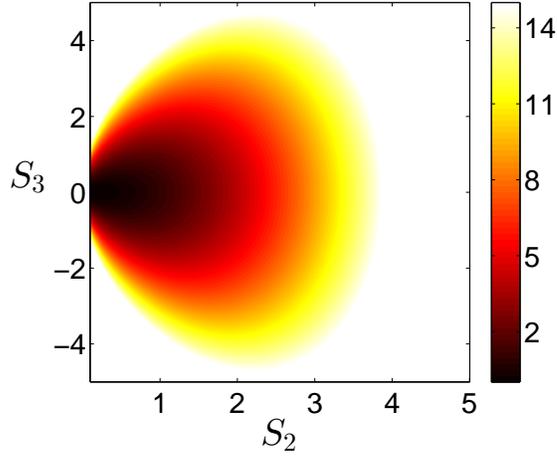}
\caption{Color map of $S_4$ (in $\mathrm{A}^4$ units as defined in Sec.~\ref{secDimAnalysis}) given by Eq.~(\ref{eqS4Trivial}) as a function of $S_2$ (in $\mathrm{A}^2$ units) and $S_3$ (in $\mathrm{A}^3$ units).}
\label{figS4Trivial}
\end{figure}

\begin{figure}
\centering
\includegraphics[width=0.5\textwidth]{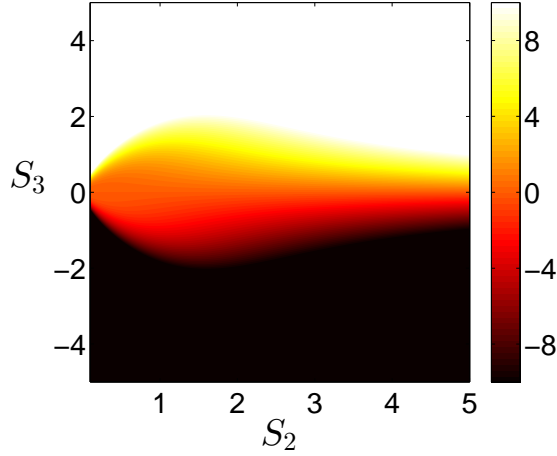}
\caption{Color map of $S_5$ (in $\mathrm{A}^5$ units as defined in Sec.~\ref{secDimAnalysis}) given by Eq.~(\ref{eqS5Trivial}) as a function of $S_2$ (in $\mathrm{A}^2$ units) and $S_3$ (in $\mathrm{A}^3$ units).}
\label{figS5Trivial}
\end{figure}

The second solution to Eqs.~(\ref{ODEs1})-(\ref{ODEs2}) corresponds to the branch $(R_0,R'_0,R''_0,R'''_0)=(1,0,2,0)$ found in Sec.~\ref{secDimAnalysis}, and is given by
\begin{eqnarray*}
R(\xi)=1+\xi^2.
\end{eqnarray*}
This leads to
\begin{eqnarray}
\label{eqS4Trivial}
& S_4=S_2^2+\frac{S_3^2}{S_2},\\
\label{eqS5Trivial}
& S_5=2S_2S_3+\frac{S_3^3}{S_2^2}.
\end{eqnarray}
These functions are plotted in Figs.~\ref{figS4Trivial} and \ref{figS5Trivial}.
Unlike the model without normal variables, we can see on Fig.~\ref{figS4Trivial} that $S_4$ does not go to 0 as $S_3$ goes to 0. As a consequence, this model allows asymmetric distribution functions and hence seems to be more physically relevant. Furthermore, one can notice the difference in the amplitude of the closures between the two models (up to two orders of magnitude) by comparing Figs.~\ref{figS4NonTrivial} and \ref{figS5NonTrivial} and Figs.~\ref{figS4Trivial} and \ref{figS5Trivial}. The Hamiltonian and the Poisson bracket resulting from this closure are given respectively by Eqs.~(\ref{hamS}) and  (\ref{brackFourFields}) with $\alpha$ and $\beta$ given by Eq.~(\ref{matAlphaBeta}) and by replacing the closures $S_4$ and $S_5$ by Eqs.~(\ref{eqS4Trivial}) and (\ref{eqS5Trivial}). 

Concerning the global Casimir invariants, we show that this Poisson bracket possesses five Casimir invariants,  as many Casimir invariants as field variables,  given by
\begin{eqnarray*}
& C_1 = \int\rho\ \mathrm{d}x,
& C_2 = \int E\ \mathrm{d}x,\\
& C_3 = \int\rho\frac{\sqrt{4S_2^3+S_3^2}}{S_2}\ \mathrm{d}x,  \qquad 
& C_4 = \int\rho\frac{S_3}{\sqrt{4S_2^3+S_3^2}}\ \mathrm{d}x,\\
& C_5 = \int\left(u+\frac{\rho}{2}\frac{S_3}{S_2}\right)\ \mathrm{d}x. 
\end{eqnarray*} 
From these expressions for the global Casimir invariants, we introduce the normal field variables $\rho$, $M=u+\rho S_3/(2S_2)$, $Q_2=\rho\sqrt{4S_2^3+S_3^2}/S_2$, $Q_3=\rho S_3/\sqrt{4S_2^3+S_3^2}$ and $E$. Consequently, Bracket~(\ref{brackFourFields}) takes the particularly simple (normal) form
\begin{equation*}
\hspace{-2.15 cm}\{F,G\}=\int\left[G_M\partial_x F_\rho-F_M\partial_x G_\rho+4\pi(G_M\widetilde{F_E}-F_M\widetilde{G_E})-2G_3\partial_x F_2-2G_2\partial_x F_3\right]\ \mathrm{d}x, 
\end{equation*}
Hamiltonian~(\ref{hamS})  becomes 
\begin{equation*}
\mathcal{H}=\frac{1}{2}\int\left(\rho M^2-M\frac{Q_3}{Q_2}+\frac{\rho}{4} Q_2^2+\frac{E^2}{4\pi}\right)\ \mathrm{d}x, 
\end{equation*}
and the Casimir invariants $C_3$, $C_4$ and $C_5$ become
\[
C_3=\int Q_2\ \mathrm{d}x,\qquad 
C_4=\int Q_3\ \mathrm{d}x,\qquad 
C_5=\int M\ \mathrm{d}x.
\]

As mentioned in Sec.~\ref{Intro}, one may be interested in using the pressure $P$ and the heat flux $q$ as dynamical variables,  instead of $S_2$ and $S_3$. Indeed, even though the reduced moments appear to be very convenient, their physical meaning may not be as clear as the usual pressure and heat flux quantities. The latter quantities can be expressed in terms of the reduced moments in the following way:
\begin{equation*}
P=\rho^3S_2=P_2-\frac{P_1^2}{P_0},  \qquad 
 q=\frac{\rho^4}{2}S_3=\frac{P_3}{2}-\frac{3}{2}\frac{P_1Px_2}{P_0}+\frac{P_1^3}{P_0^2}, 
\end{equation*}
in terms of  which  the closures take the form 
\[
  S_4= \frac{1}{\rho^5}\left(\frac{P^2}{\rho}+\frac{4q^2}{P} \right), \qquad 
  S_5 = \frac{4q}{\rho^6}\left(\frac{P}{\rho}+\frac{2q^2}{P^2} \right).
\]
Expressed in terms of these variables, Bracket~(\ref{brackFourFields}) becomes
\begin{eqnarray*}
&{\ }& \hspace{-1.75 cm}\{F,G\}=\int\Bigg[G_u\partial_x F_\rho-F_u\partial_x G_\rho+\frac{3P}{\rho}(G_u\partial_x F_P-F_u\partial_x G_P)\nonumber\\
&+&4\pi(G_u\widetilde{F_E}-F_u\widetilde{G_E})+\frac{2}{\rho}(G_u F_P-F_u G_P)\partial_x P+\frac{4q}{\rho}(G_u\partial_x F_q-F_u\partial_x G_q)\nonumber\\
&+&\frac{3}{\rho}(G_u F_q-F_u G_q)\partial_x q+\rho^4\bar{\alpha}_{22}F_PG_P+\rho^5\bar{\alpha}_{23}F_PG_q+\rho^5\bar{\alpha}_{32}F_qG_P\nonumber\\
&+&\rho^6\bar{\alpha}_{33}F_qG_q+\rho^2\bar{\beta}_{22}G_P\partial_x(\rho^2F_P)+\rho^3\bar{\beta}_{23}G_q\partial_x(\rho^2F_P)+\rho^2\bar{\beta}_{32}G_P\partial_x(\rho^3F_q)\nonumber\\
&+&\rho^3\bar{\beta}_{33}G_q\partial_x(\rho^3F_q)\Bigg]\ \mathrm{d}x,
\end{eqnarray*}
where
$$
\bar{\alpha}=\partial_x\left(
\begin{array}{cc}
4q/\rho^4 & 4q^2/(\rho^5P)-P^2/(2\rho^6)\\
6q^2/(\rho^5P)-3P^2/(2\rho^6) & 6q^3/(\rho^6P^2) - 3Pq/\rho^7
\end{array}
\right),
$$
and
$$
\bar{\beta}=\left(
\begin{array}{cc}
8q/\rho^4 & 10q^2/(\rho^5P)-2P^2/\rho^6\\
10q^2/(\rho^5P)-2P^2/\rho^6 & 12q^3/(\rho^6P^2) - 6Pq/\rho^7
\end{array}
\right).
$$
Hamiltonian~(\ref{hamS}) takes the simple form
\begin{equation*}
\mathcal{H}=\frac{1}{2}\int\left(\rho u^2+P+\frac{E^2}{4\pi}\right)\ \mathrm{d}x, 
\end{equation*}
and the equations of motion,  obtained from $\partial_t F=\{F,{\cal H}\}$, are 
\begin{eqnarray*}
\partial_t\rho&=-\rho\partial_x u-u\partial_x\rho,\\
\partial_tu&=-u\partial_xu-\frac{1}{\rho}\partial_x P-\widetilde{E},\\
\partial_t P&=-u\partial_x P-3P\partial_xu-2\partial_x q,\\
\partial_t q&=-u\partial_x q-4q\partial_xu-2\partial_x\left(\frac{q^2}{P}\right)+\frac{1}{4\rho^3}\partial_x\left(\rho^2P^2\right),\\
\partial_t E&=4\pi\widetilde{\rho u}.
\end{eqnarray*}
We notice that these equations are identical (at least the ones concerning $\rho$, $u$, $P$ and $q$) to the equations obtained with a bi-delta reduction~\cite{Fox09,Yuan11,Chalons12,Cheng14}. Therefore, as a by-product of our reduction procedure, we have proved here that the bi-delta reduction is Hamiltonian.  This can also be shown  by effecting a chain rule calculation relating the Vlasov-Poisson bracket \cite{Morrison80b} to that of fluid streams \cite{hagstrom}.

A benefit of knowing the Hamiltonian structure of the reduced model is the ability to use the Poisson bracket to  obtain the additional invariants, e.g., Casimir invariants, that  can be tricky to derive directly from  the equations of motion. For example,  the global Casimir invariants $C_3$, $C_4$ and $C_5$ for the present system are seen to be 
\begin{equation*}
\hspace*{-1.75 cm} C_3=\int\sqrt{\frac{P}{\rho}+\frac{q^2}{P^2}}\ \mathrm{d}x,\quad
C_4=\int\!\rho\, q\sqrt{\frac{\rho}{P^3+\rho q^2}}\ \mathrm{d}x,\quad 
C_5=\int\left(u+\frac{q}{P}\right)\ \mathrm{d}x.
\end{equation*}
We note that these invariants can be used to check the validity of  numerical algorithms  used for the integration of the equations of motion. 

\subsection{Comparison with Hamiltonian fluid models with 3+1 fields}

The same analysis done for 4+1 fields can be carried out for fluid models with 3+1 fields, namely with the field variables $(P_0,P_1,P_2,E)$ or equivalently $(\rho, u,S_2,E)$. This was  partly done in Ref.~\cite{Perin14} (in the absence of electric field), where it was shown that Hamiltonian fluid models are given by closures $S_3$ that only depend on $S_2$. This is also evident from  \ref{appendixConstraints},  where the conditions given by Eq.~(\ref{eq:gammagamma}) are automatically satisfied (since there is only one value for the indices). The fact that the closure for 3+1 Hamiltonian fluid models only depends  on $S_2$ is similar to the fact that the closures for 4+1 fluid models are given by $S_4$ and $S_5$ as functions of only $S_2$ and $S_3$. 

A specific closure $S_3(S_2)$, which  corresponds to the dimensional analysis performed in the present work, is given by
$$
S_3=\lambda S_2^{3/2},
$$
where $\lambda$ is a dimensionless constant. The Poisson bracket for this closure is
\begin{eqnarray*}
\hspace{-1.0 cm}\{F,G\}_3&=&\int \Bigg\{G_u \partial_x F_\rho -F_u\partial_x G_\rho +4\pi(G_u\widetilde{F_E}-F_u\widetilde{G_E})
\\
&-& \frac{1}{\rho}(G_u F_2-F_u G_2)\partial_x S_2 
+2\lambda S_2^{3/2}\left[ \frac{G_2}{\rho}\partial_x\left(\frac{F_2}{\rho}\right)- \frac{F_2}{\rho} \partial_x\left(\frac{G_2}{\rho}\right)\right]\Bigg\}\ {\rm d}x.
\end{eqnarray*}
It should be noted that the dimensional analysis provides a family of models (labeled by $\lambda$). However there are only three fundamentally different models:  one for $\lambda=0$ and the others for  $\lambda=\pm1$, since all of the other models can be rescaled to $\lambda=\pm1$  by an appropriate rescaling of $S_2$, e.g., $\bar{S_2}=S_2/\lambda^2$. Moreover, the two models $\lambda=1$ and $\lambda=-1$ are linked by symmetry \cite{Perin14}. The model for $\lambda=0$ has the two following global Casimir invariants: 
\[
 C_1 = \int \rho\ {\rm d}x \qquad \mathrm{and} \qquad  C_2 = \int E\ {\rm d} x,
\]
in addition to the family of Casimir invariants
$$
C=\int \rho \kappa(S_2)\ {\rm d}x,
$$
for any scalar function $\kappa$. The two Casimir invariants $C_1$ and $C_2$ are identical to the ones for 4+1 fields. Concerning the model with $\lambda =1$, the Poisson bracket with 3+1 fields has $C_1$ and $C_2$ as Casimir invariants, and also has two additional Casimir invariants
\[
  C_3 = \int \rho S_2^{1/4} {\rm d} x  \qquad \mathrm{and} \qquad 
  C_4 = \int \left(u -2\rho S_2^{1/2} \right) {\rm d}x.
\]
Therefore, in total  it has four Casimir invariants, i.e., as many as the number of field variables. 
The common point between this 3+1 model with $\lambda=1$ and the 4+1 field model with normal field  variables is that both have  a generalized velocity as Casimir invariant. It should also be noticed that the 3+1 fluid model has one Casimir invariant of the entropy type, i.e., of the form $\int \rho \phi(S_2,\ldots,S_N)\ {\rm d} x$, whereas the 4+1 fluid model has two of this type.

\section{Summary}

In summary, starting from the one-dimensional Vlasov-Amp\`ere equations, we built two Hamiltonian models with  the first four moments of the Vlasov distribution function and the electric field as dynamical variables. Our reduction method relied  on the preservation of the Hamiltonian structure of the Vlasov-Amp\`ere model. The closures we obtain were derived from a dimensional analysis argument. We showed that there are only two Hamiltonian closures obtained by this method. A fundamental difference between these two models was  characterized by their Casimir invariants: one model has only two global Casimir invariants (preserved from the Vlasov-Amp\`ere system), whereas the second model has three additional ones, two of the entropy-type and one generalized velocity.

\section*{Acknowledgments}
This work was supported by the Agence Nationale de la Recherche (ANR GYPSI) and by the European Community under the contract of Association between EURATOM, CEA, and the French Research Federation for fusion study. The views and opinions expressed herein do not necessarily reflect those of the European Commission. ET acknowledges also the financial support from the CNRS through the PEPS project GEOPLASMA2.  PJM was  supported by U.S. Dept.\ of Energy Contract \# DE-FG02-04ER54742.

\appendix

\section{Independence of $\alpha$ and $\beta$ of Bracket~(\ref{brackFourFields}) on $\rho,u,$ and $E$}
\label{appendixJacobi}

In this appendix, we consider the following bracket defined  on functionals of the form $F[\rho,u,S_2,\dots,S_N,E]$ for $N\geq2$:
\begin{eqnarray}
\label{brackAlphaBeta}
\hspace{-1 cm} \{F,G\}&=&\int\Bigg[G_u\partial_x F_\rho-F_u\partial_x G_\rho+4\pi(G_u\widetilde{F_E}-F_u\widetilde{G_E})-\frac{1}{\rho}(G_u F_i-F_u G_i)\partial_x S_i\nonumber\\
& & \hspace{3 cm}  +\ \alpha_{ij}\frac{F_i}{\rho}\frac{G_j}{\rho}+\partial_x\left(\frac{F_i}{\rho}\right)\beta_{ij}\frac{G_j}{\rho}\Bigg]\ \mathrm{d}x,
\end{eqnarray}
where $\alpha$ and $\beta$ are matrices satisfying $\beta^t=\beta$ and $\partial_x\beta=\alpha+\alpha^t$, assuring  antisymmetry of the bracket. Here we assume a priori that $\alpha$ and $\beta$ depend on both the dynamical variables $\rho$, $u$, $S_k$ (for $k\geq 2$) and $E$, and their derivatives $\partial_x^n \rho$, $\partial_x^n u$, $\partial_x^n S_k$ and $\partial_x^n E$ for $n\geq 1$.  Repeated indices are implicitly summed from 2 to $N$, unless specified. We seek  necessary conditions on $\alpha$ and $\beta$ for Bracket~(\ref{brackAlphaBeta}) to satisfy the Jacobi identity, 
$$
\{F,\{G,H\}\}+\{H,\{F,G\}\}+\{G,\{H,F\}\}=0.
$$
In this appendix, we prove that $\alpha$ and $\beta$ do not depend on the variables $\rho$, $u$ and $E$ and their derivatives $\partial_x^n \rho$, $\partial_x^n u$ and $\partial_x^n E$ for $n\geq 1$. 

First we split Bracket~(\ref{brackAlphaBeta}) into two parts
$$
\{F,G\}=\{F,G\}^J+\{F,G\}^*,
$$
where the first part,
$$
\{F,G\}^J=\int\left[G_u\partial_x F_\rho-F_u\partial_x G_\rho+4\pi(G_u\widetilde{F_E}-F_u\widetilde{G_E})-\frac{1}{\rho}( G_u F_i-F_u G_i)\partial_x S_i\right]\ \mathrm{d}x,
$$
satisfies the Jacobi identity~\cite{Morrison82,Morrison98}. The Jacobi identity is then equivalent to
\begin{equation}
\label{Jacobi}
\{F,\{G,H\}^J\}^*+\{F,\{G,H\}^*\}^J+\{F,\{G,H\}^*\}^*+\circlearrowleft_{(F,G,H)}=0,
\end{equation}
where $\circlearrowleft_{(F,G,H)}$ denotes the summation over circular permutation of any  three functionals $F$, $G$ and $H$. Using the lemma stating that only the functional derivatives with respect to the explicit dependence on the variables need  be taken into account for the Jacobi identity \cite{Morrison82}, the first term becomes 
\begin{eqnarray}
\hspace*{-1.5 cm}\{F,\{G,H\}^J\}^* =&\int\Bigg[\frac{\alpha_{ij}}{\rho}\frac{F_i}{\rho}\left[H_u\partial_x\left(\frac{G_j}{\rho}\right)-G_u\partial_x\left(\frac{H_j}{\rho}\right)+\frac{G_j}{\rho}\partial_x H_u-\frac{H_j}{\rho}\partial_x G_u\right]\nonumber\\
\label{eq:Jstar}
&+\frac{\beta_{ij}}{\rho}\partial_x\left(\frac{F_i}{\rho}\right)\left(\frac{G_j}{\rho}\partial_x H_u-\frac{H_j}{\rho}\partial_x G_u\right)\Bigg]\ \mathrm{d}x,
\end{eqnarray}
where we have used the fact that $\beta$ is symmetric. The second term in Eq.~(\ref{Jacobi}) is 
\begin{eqnarray}
\hspace*{-2.0 cm}\{F,\{G,H\}^*\}^J&=&\int\Bigg[\{G,H\}^*_u\partial_x F_\rho+\{G,H\}^*_\rho\partial_x F_u+4\pi(\{G,H\}^*_u\widetilde{F_E}-\{G,H\}^*_E\widetilde{F_u})\nonumber\\
& & -\ \frac{1}{\rho}(\{G,H\}^*_u F_i-F_u \{G,H\}^*_i)\partial_x S_i\Bigg]\ \mathrm{d}x,\label{eq:starJ}
\end{eqnarray}
where
\begin{eqnarray*}
\{G,H\}^*_\rho &=&\left(\frac{\alpha_{ji}}{\rho}-\frac{\alpha_{ij}}{\rho}\right)\frac{G_i}{\rho}\frac{H_j}{\rho}+\frac{\beta_{ij}}{\rho}\left[\partial_x\left(\frac{H_j}{\rho}\right)\frac{G_i}{\rho}-\partial_x\left(\frac{G_i}{\rho}\right)\frac{H_j}{\rho}\right]\\
&& +\ (-1)^n\partial_x^n\left(\left[\frac{\partial\alpha_{ij}}{\partial\partial_x^n\rho}\frac{G_i}{\rho}+\frac{\partial\beta_{ij}}{\partial\partial_x^n\rho}\partial_x\left(\frac{G_i}{\rho}\right)\right]\frac{H_j}{\rho}\right)\\
\{G,H\}^*_u &=&(-1)^n\partial_x^n\left(\left[\frac{\partial\alpha_{ij}}{\partial\partial_x^nu}\frac{G_i}{\rho}+\frac{\partial\beta_{ij}}{\partial\partial_x^nu}\partial_x\left(\frac{G_i}{\rho}\right)\right]\frac{H_j}{\rho}\right),\\
\{G,H\}^*_k &=&(-1)^n\partial_x^n\left(\left[\frac{\partial\alpha_{ij}}{\partial\partial_x^nS_k}\frac{G_i}{\rho}+\frac{\partial\beta_{ij}}{\partial\partial_x^nS_k}\partial_x\left(\frac{G_i}{\rho}\right)\right]\frac{H_j}{\rho}\right),\\
\{G,H\}^*_E &=&(-1)^n\partial_x^n\left(\left[\frac{\partial\alpha_{ij}}{\partial\partial_x^nE}\frac{G_i}{\rho}+\frac{\partial\beta_{ij}}{\partial\partial_x^nE}\partial_x\left(\frac{G_i}{\rho}\right)\right]\frac{H_j}{\rho}\right).
\end{eqnarray*}
We consider the terms of the type $(F_u,G_i,H_j)$ in the Jacobi identity~(\ref{Jacobi}). These terms only come from Eqs.~(\ref{eq:Jstar}) and (\ref{eq:starJ}). By using successive integrations by parts and assuming that the boundary conditions are such that the associated boundary integrals vanish, the Jacobi identity for these terms becomes
\begin{eqnarray}
\int\Bigg\{\left[\frac{\partial\alpha_{ij}}{\partial\partial_x^n\rho}\frac{G_i}{\rho}+\frac{\partial\beta_{ij}}{\partial\partial_x^n\rho}\partial_x\left(\frac{G_i}{\rho}\right)\right]\frac{H_j}{\rho}\partial_x^{n+1}F_u-\frac{H_j}{\rho}\partial_x\left(\alpha_{ij}\frac{G_i}{\rho}\frac{F_u}{\rho}\right)\nonumber\\
\label{tmp8}
-\alpha_{ji}\partial_x\left(\frac{G_i}{\rho}\right)\frac{H_j}{\rho}\frac{F_u}{\rho}-4\pi\left[\frac{\partial\alpha_{ij}}{\partial\partial_x^nE}\frac{G_i}{\rho}+\frac{\partial\beta_{ij}}{\partial\partial_x^nE}\partial_x\left(\frac{G_i}{\rho}\right)\right]\frac{H_j}{\rho}\partial_x^n\widetilde{F_u}\\
+\left[\frac{\partial\alpha_{ij}}{\partial\partial_x^nS_k}\frac{G_i}{\rho}+\frac{\partial\beta_{ij}}{\partial\partial_x^nS_k}\partial_x\left(\frac{G_i}{\rho}\right)\right]\frac{H_j}{\rho}\partial_x^n\left(\frac{F_u}{\rho} \partial_x S_k\right)\Bigg\}\ \mathrm{d}x+\circlearrowleft_{(F,G,H)}=0.\nonumber
\end{eqnarray}
Choosing $F=\int u\ \mathrm{d}x$, $G=\int\rho S_l\ \mathrm{d}x$ and $H=\rho S_m$, Eq.~(\ref{tmp8}) leads to the necessary condition
\begin{equation}
\label{tmp}
\frac{1}{\rho}\partial_x\alpha_{lm}=\frac{\alpha_{lm}}{\rho^2}\partial_x\rho+\frac{\partial\alpha_{lm}}{\partial\partial_x^nS_k}\partial_x^n\left(\frac{1}{\rho} \partial_x S_k\right).
\end{equation}
However, we have by definition
$$
\partial_x\alpha_{lm}=\frac{\partial\alpha_{lm}}{\partial\partial_x^n\rho}\partial_x^{n+1}\rho+\frac{\partial\alpha_{lm}}{\partial\partial_x^nu}\partial_x^{n+1}u+\frac{\partial\alpha_{lm}}{\partial\partial_x^nS_k}\partial_x^{n+1}S_k+\frac{\partial\alpha_{lm}}{\partial\partial_x^nE}\partial_x^{n+1}E+\frac{\partial\alpha_{lm}}{\partial x},
$$
where the summation over $n$ is implicit and $\partial\alpha_{lm}/\partial x$ denotes the derivative of $\alpha_{lm}$ with respect to its explicit dependence on $x$. Eventually, Eq.~(\ref{tmp}) writes
\begin{eqnarray}
\frac{1}{\rho}\left(\frac{\partial\alpha_{lm}}{\partial\partial_x^n\rho}\partial_x^{n+1}\rho+\frac{\partial\alpha_{lm}}{\partial\partial_x^nu}\partial_x^{n+1}u+\frac{\partial\alpha_{lm}}{\partial\partial_x^nS_k}\partial_x^{n+1}S_k+\frac{\partial\alpha_{lm}}{\partial\partial_x^nE}\partial_x^{n+1}E+\frac{\partial\alpha_{lm}}{\partial x}\right)\nonumber\\
\label{tmpBis}
=\frac{\alpha_{lm}}{\rho^2}\partial_x\rho+\frac{\partial\alpha_{lm}}{\partial\partial_x^nS_k}\partial_x^n\left(\frac{1}{\rho} \partial_x S_k\right).
\end{eqnarray}
By canceling the only term that depend on $\partial_x^{\nu+1}\rho$ in Eq.~(\ref{tmpBis}), we can show that
\begin{equation*}
\frac{\partial\alpha_{lm}}{\partial\partial_x^\nu \rho}=0.
\end{equation*}
By performing an induction on $\nu$ down to $\nu=0$, we can show that $\alpha$ does not depend on $\rho$ and its derivatives. Because the dynamical variables are independent, using the same reasoning we prove that $\alpha$ cannot depend on $v$, $E$ and their derivatives, nor can  it depend explicitly on $x$. The same result can be obtained for $\beta$ by choosing $G=\int\rho S_lx\ \mathrm{d}x$. Therefore a necessary (however not sufficient) condition for Bracket~(\ref{brackAlphaBeta}) to satisfy the Jacobi identity is that $\alpha$ and $\beta$  do not  depend explicitly on $x$, $\rho$, $u$ and $E$, as well as the derivatives $\partial_x^n \rho$, $\partial_x^n u$ and $\partial_x^n E$ for all $n\in \mathbb{N}$.

\section{Dependence  of $\alpha$ and $\beta$ of  Bracket~(\ref{brackFourFields}) on $S_k$}
\label{appendixDerivatives}

In this appendix, we derive some necessary conditions on the dependence of $\alpha$ and $\beta$  (and their derivatives) of Bracket~(\ref{brackFourFields}) on $S_k$. Following \ref{appendixJacobi}, we consider two sets of functionals
$$
(F,G,H)=\left(\int u x\ \mathrm{d}x, \int\rho S_l\ \mathrm{d}x,\rho S_m\right),
$$
and
$$
(F,G,H)=\left(\int u x\ \mathrm{d}x,\int\rho S_l x\ \mathrm{d}x,\rho S_m\right),
$$
which we insert into  Eq.~(\ref{tmp8}).  Thus we find the necessary conditions
\begin{equation}
\label{eqAlpha}
\alpha_{lm}=n\frac{\partial\alpha_{lm}}{\partial\partial_x^nS_k}\partial_x^nS_k,\qquad n\frac{\partial\beta_{lm}}{\partial\partial_x^nS_k}\partial_x^nS_k=0,
\end{equation}
for all $l$ and $m$,  where we recall the implicit summation over repeated indices. We assume that $\alpha$ and $\beta$ depend on the derivatives of $S_k$ up to order $\nu$, where 
$$
\nu=\max\{n\in \mathbb{N} \mbox{ s.t. } \partial\alpha/\partial\partial_x^nS\neq0\ \mbox{ or }\ \partial\beta/\partial\partial_x^nS\neq0\}.
$$
From the first of Eqs.~(\ref{eqAlpha}) we have
\begin{equation}
\label{rnd99}
\hspace*{-1.25 cm}\partial_x\beta_{lm}=\sum\limits_{n=0}^\nu\frac{\partial\beta_{lm}}{\partial\partial_x^nS_k}\partial_x^{n+1}S_k=\alpha_{lm}+\alpha_{ml}=\sum\limits_{n=0}^\nu n\left[\frac{\partial\alpha_{lm}}{\partial\partial_x^nS_k}+\frac{\partial\alpha_{ml}}{\partial\partial_x^nS_k}\right]\partial_x^nS_k.
\end{equation}
Differentiating Eq.~(\ref{rnd99}) with respect to $\partial^{\nu+1}S_j$ leads to
\begin{equation*}
\frac{\partial\beta_{lm}}{\partial\partial_x^\nu S_j}=0.
\end{equation*}
As a consequence, the highest derivatives of $S$ appear in $\alpha$; thus $\nu$ becomes
$$
\nu=\max\{n\in \mathbb{N} \mbox{ s.t. }\partial\alpha/\partial\partial_x^nS\neq0\}.
$$
The Jacobi identity~(\ref{Jacobi}) reduces to:
\begin{eqnarray}
\hspace*{-2.5 cm}\{F,\{G,H\}\}+\circlearrowleft_{(F,G,H)}&=&\int\!\left[\alpha_{ij}\frac{F_i}{\rho}\frac{\{G,H\}^*_j}{\rho}+\beta_{ij}\frac{\{G,H\}^*_j}{\rho}\partial_x\left(\frac{F_i}{\rho}\right)\right]\ \mathrm{d}x\label{eq:Jbrackstar}
\\
&& \hspace{7cm}+\circlearrowleft_{(F,G,H)} = 0. 
\nonumber
\end{eqnarray}
This identity corresponds to the Jacobi identity for the subalgebra of observables $F[\rho,S_2,\dots,S_N]$. Expanding Eq.~(\ref{eq:Jbrackstar}) gives
\begin{eqnarray*}
\hspace*{-2.0 cm}\{F,\{G,H\}\}&+&\circlearrowleft_{(F,G,H)}=\int\Bigg\{\frac{\alpha_{ij}}{\rho}\frac{F_i}{\rho}(-1)^n\partial_x^n\left(\left[\frac{\partial\alpha_{kl}}{\partial\partial_x^nS_j}\frac{G_k}{\rho}+\frac{\partial\beta_{kl}}{\partial\partial_x^nS_j}\partial_x\left(\frac{G_k}{\rho}\right)\right]\frac{H_l}{\rho}\right)\\
&+&(-1)^{n+1}\frac{F_i}{\rho}\partial_x\left[\frac{\beta_{ij}}{\rho}\partial_x^n\left(\left[\frac{\partial\alpha_{kl}}{\partial\partial_x^nS_j}\frac{G_k}{\rho}+\frac{\partial\beta_{kl}}{\partial\partial_x^nS_j}\partial_x\left(\frac{G_k}{\rho}\right)\right]\frac{H_l}{\rho}\right)\right]\\
&+&\partial_x^n\left[\frac{\alpha_{lj}}{\rho}\frac{H_l}{\rho}+\frac{\beta_{lj}}{\rho}\partial_x\left(\frac{H_l}{\rho}\right)\right]\frac{\partial\alpha_{ik}}{\partial\partial_x^nS_j}\frac{F_i}{\rho}\ \frac{G_k}{\rho}\\
&-&\frac{F_i}{\rho}\partial_x\left(\partial_x^n\left[\frac{\alpha_{lj}}{\rho}\frac{H_l}{\rho}+\frac{\beta_{lj}}{\rho}\partial_x\left(\frac{H_l}{\rho}\right)\right]\frac{\partial\beta_{ik}}{\partial\partial_x^nS_j}\frac{G_k}{\rho}\right)\\
&+&\partial_x^n\left[\frac{\alpha_{kj}}{\rho}\frac{G_k}{\rho}+\frac{\beta_{kj}}{\rho}\partial_x\left(\frac{G_k}{\rho}\right)\right]\left[\frac{\partial\alpha_{li}}{\partial\partial_x^nS_j}\frac{H_l}{\rho}+\frac{\partial\beta_{li}}{\partial\partial_x^nS_j}\partial_x\left(\frac{H_l}{\rho}\right)\right]\frac{F_i}{\rho}\Bigg\}\ \mathrm{d}x.
\end{eqnarray*}
Choosing consecutively 
\begin{equation*}
(F,G,H)=\left(\rho S_i,\int\rho S_k\ \mathrm{d}x,\int\rho S_l\ \mathrm{d}x\right),
\end{equation*}
\begin{equation*}
(F,G,H)=\left(\rho S_i,\int\rho S_k x\ \mathrm{d}x,\int\rho S_l\ \mathrm{d}x\right),
\end{equation*}
and
\begin{equation*}
(F,G,H)=\left(\rho S_i,\int\rho S_k\ \mathrm{d}x,\int\rho S_l x\ \mathrm{d}x\right),
\end{equation*}
we get the following three conditions:
\begin{eqnarray}
\label{rnd54}
&\left[\frac{\partial\alpha_{ik}}{\partial\partial_x^\nu S_j}-\frac{\partial\beta_{ik}}{\partial\partial_x^{\nu-1}S_j}\right]\alpha_{lj}+\alpha_{kj}\frac{\partial\alpha_{li}}{\partial\partial_x^\nu S_j}=0,\\
\label{rnd55}
&\beta_{kj}\frac{\partial\alpha_{li}}{\partial\partial_x^\nu S_j}=0,\\
\label{rnd56}
&\beta_{lj}\frac{\partial\beta_{ik}}{\partial\partial_x^{\nu-1}S_j}=0.
\end{eqnarray}
Due to the fact that $\partial_x\beta=\alpha+\alpha^t$, Eqs.~(\ref{rnd55}) and (\ref{rnd56}) are redundant. We assume that $\alpha$ depends only linearly on $\partial_x^\nu S_k$. We show in \ref{appendixConstraints} that this is the case for fluid brackets. As a consequence, we write
$$
\alpha_{lj}(S,\partial_x S,\dots,\partial_x^\nu S)=A_{lj}(S,\partial_x S,\dots,\partial_x^{\nu-1}S)+\gamma_{ljm}(S,\partial_x S,\dots,\partial_x^{\nu-1}S)\partial_x^\nu S_m.
$$
By inserting this expression into Eq.~(\ref{rnd54}),  for the Jacobi identity we need to impose
$$
\gamma_{ljm}\left[\gamma_{ikj}-\frac{\partial\beta_{ik}}{\partial\partial_x^{\nu-1}S_j}\right]+\gamma_{kjm}\gamma_{lij}=0,
$$
for all $(i,k,l,m)$ to make the term proportional to $\partial_x^\nu S_m$ vanish. However, thanks to Eq.~(\ref{rnd99}) we have
$$
\frac{\partial\beta_{ik}}{\partial\partial_x^{\nu-1}S_j}=\gamma_{ikj}+\gamma_{kij}.
$$
This eventually leads to the following conditions: 
\begin{equation}
\gamma_{ljm}\gamma_{kij}=\gamma_{kjm}\gamma_{lij}.
\label{eq:gammagamma}
\end{equation}
These commutation relations remind us of the conditions for  Lie-Poisson brackets based on Lie algebra extensions to satisfy the Jacobi identity of  Ref.~\cite{Thiffeault00}. These conditions on the tensor $\gamma$ are necessary but not sufficient. 

\section{Jacobi identity for fluid models}
\label{appendixConstraints}

In this appendix we find necessary and sufficient conditions for  the Jacobi identity for Bracket~(\ref{brackFourFields}). 
We start from the one-dimensional Vlasov-Amp\`ere bracket given by Eq.~(\ref{brackVP}) and perform a change of variables, from $f$ to $(\rho,u, S_{n\geq 2})$ defined by
\begin{equation*}
\rho = \int f\ {\rm d}v,\qquad
\rho u = \int vf\ {\rm d} v,\qquad
\rho^{n+1}S_n = \int \left(v-u\right)^n f\ {\rm d}v.
\end{equation*} 
Using the following chain rule expression for the functional derivative with respect to $f$,
$$
F_f=\bar{F}_\rho+\frac{v-u}{\rho}\bar{F}_u+\bar{F}_n\left[\frac{(v-u)^n}{\rho}-\frac{n+1}{\rho}S_n-n\frac{S_{n-1}}{\rho}\frac{(v-u)}{\rho}\right],
$$
and after some algebra, we show that the Poisson bracket~(\ref{brackVP}) reduces to Eq.~(\ref{brackAlphaBeta}) with $\alpha$ and $\beta$ given by
\begin{eqnarray}
&& \hspace{-1 cm} \alpha_{nm}=n\partial_x S_{n+m-1}-nS_{n-1}\partial_x S_m-n(m+1)S_m\partial_x S_{n-1} 
 -nmS_{m-1}\partial_x S_n,\label{eq:fluidA}\\
\label{eq:fluidB}
&& \hspace{-1 cm} \beta_{nm}=(m+n)S_{n+m-1}-m(n+1)S_nS_{m-1}-n(m+1)S_mS_{n-1},
\end{eqnarray}
where $n,m\geq 2$ and $S_1=0$. The resulting bracket is of Poisson type. Next,
we truncate the matrices $\alpha$ and $\beta$ such that $\alpha_{mn}=0$ and $\beta_{mn}=0$ for $m> N$ or $n> N$. The matrices $\alpha$ and $\beta$ depend on $S_n$ for $n=2,\ldots,2N-1$. We restrict ourselves to the case where $\alpha$ and $\beta$ are functions of $(S_2,\ldots,S_N)$, i.e., we introduce $N-1$ closures $S_k=S_k(S_2,\ldots,S_N)$ for $k=N+1,\ldots,2N-1$. In this truncation/reduction, the bracket is no longer of Poisson type in general. In this appendix, we establish the necessary and sufficient conditions for the Jacobi identity to be satisfied. From  \ref{appendixJacobi} and \ref{appendixDerivatives}, this Jacobi identity is seen to be 
\begin{eqnarray}
\hspace*{-1.5 cm}\{F,\{G,H\}\}&&+\circlearrowleft_{(F,G,H)}=\int\Bigg\{\partial_x\left(\frac{F_i}{\rho}\right)\frac{1}{\rho}\partial_x\left(\frac{G_k}{\rho}\right)\frac{H_l}{\rho}\Bigg[\beta_{ij}\frac{\partial\beta_{kl}}{\partial S_j}-\beta_{ij}\frac{\partial\alpha_{kl}}{\partial\partial_x S_j}\nonumber\\
&&-\beta_{kj}\frac{\partial\alpha_{li}}{\partial\partial_x S_j}\Bigg]+\frac{F_i}{\rho}\frac{1}{\rho}\frac{G_k}{\rho}\frac{H_l}{\rho}\Bigg[\alpha_{ij}\frac{\partial\alpha_{kl}}{\partial S_j}+\alpha_{lj}\frac{\partial\alpha_{ik}}{\partial S_j}+\alpha_{kj}\frac{\partial\alpha_{li}}{\partial S_j}\nonumber\\
&&-\alpha_{ij}\partial_x\left(\frac{\partial\alpha_{kl}}{\partial\partial_x S_j}\right)-\alpha_{lj}\partial_x\left(\frac{\partial\alpha_{ik}}{\partial\partial_x S_j}\right)-\alpha_{kj}\partial_x\left(\frac{\partial\alpha_{li}}{\partial\partial_x S_j}\right)\Bigg]\nonumber\\
&&+\frac{F_i}{\rho}\frac{1}{\rho}\partial_x\left(\frac{G_k}{\rho}\right)\frac{H_l}{\rho}\Bigg[\alpha_{ij}\frac{\partial\beta_{kl}}{\partial S_j}+\beta_{kj}\frac{\partial\alpha_{li}}{\partial S_j}-\alpha_{ij}\frac{\partial\alpha_{kl}}{\partial\partial_x S_j}-\alpha_{lj}\frac{\partial\alpha_{ik}}{\partial\partial_x S_j}\nonumber\\
&&-\beta_{kj}\partial_x\left(\frac{\partial\alpha_{li}}{\partial\partial_x S_j}\right)\Bigg]+\partial_x\left(\frac{F_i}{\rho}\right)\frac{1}{\rho}\frac{G_k}{\rho}\frac{H_l}{\rho}\Bigg[\beta_{ij}\frac{\partial\alpha_{kl}}{\partial S_j}+\alpha_{lj}\frac{\partial\beta_{ik}}{\partial S_j}\nonumber\\
&&-\beta_{ij}\partial_x\left(\frac{\partial\alpha_{kl}}{\partial\partial_x S_j}\right)-\alpha_{lj}\frac{\partial\alpha_{ik}}{\partial\partial_x S_j}-\alpha_{kj}\frac{\partial\alpha_{li}}{\partial\partial_x S_j}\Bigg]\nonumber\\
&&+\frac{1}{\rho}\partial_x\left(\frac{H_l}{\rho}\right)\frac{F_i}{\rho}\frac{G_k}{\rho}\Bigg[\beta_{lj}\frac{\partial\alpha_{ik}}{\partial S_j}-\alpha_{ij}\frac{\partial\alpha_{kl}}{\partial\partial_x S_j}+\alpha_{kj}\frac{\partial\beta_{li}}{\partial S_j}\nonumber\\
&&-\beta_{lj}\partial_x\left(\frac{\partial\alpha_{ik}}{\partial\partial_x S_j}\right)-\alpha_{kj}\frac{\partial\alpha_{li}}{\partial\partial_x S_j}\Bigg]+\partial_x\left(\frac{F_i}{\rho}\right)\frac{1}{\rho}\partial_x\left(\frac{H_l}{\rho}\right)\frac{G_k}{\rho}\Bigg[\nonumber\\
\label{rnd7}
&&\beta_{lj}\frac{\partial\beta_{ik}}{\partial S_j}-\beta_{ij}\frac{\partial\alpha_{kl}}{\partial\partial_x S_j}-\beta_{lj}\frac{\partial\alpha_{ik}}{\partial\partial_x S_j}\Bigg]\\
&&+\frac{1}{\rho}\partial_x\left(\frac{G_k}{\rho}\right)\partial_x\left(\frac{H_l}{\rho}\right)\frac{F_i}{\rho}\left[\beta_{kj}\frac{\partial\beta_{li}}{\partial S_j}-\beta_{lj}\frac{\partial\alpha_{ik}}{\partial\partial_x S_j}-\beta_{kj}\frac{\partial\alpha_{li}}{\partial\partial_x S_j}\right]\Bigg\}\ \mathrm{d}x.\nonumber
\end{eqnarray}
Choosing $F=\rho S_i$, $G=\int\rho S_k\ \mathrm{d}x$ and $H=\int\rho S_l\ \mathrm{d}x$, Eq.~(\ref{rnd7}) reduces to
\begin{eqnarray}
&& \frac{\alpha_{ij}}{\rho}\bigg(\frac{\partial\alpha_{lk}}{\partial S_j}-\partial_x \frac{\partial\alpha_{lk}}{\partial\partial_x S_j}\bigg)+\partial_x\left(\frac{\beta_{ij}}{\rho}\left[\partial_x\left(\frac{\partial\alpha_{lk}}{\partial\partial_x S_j}\right)-\frac{\partial\alpha_{lk}}{\partial S_j}\right]-\frac{\alpha_{kj}}{\rho}\frac{\partial\beta_{il}}{\partial S_j}\right) \nonumber \\
\label{eqSplit}
&& +\frac{\alpha_{kj}}{\rho}\frac{\partial\alpha_{il}}{\partial S_j}+\frac{1}{\rho}\frac{\partial\alpha_{il}}{\partial\partial_x S_j}\partial_x\alpha_{kj}+\frac{\alpha_{lj}}{\rho}\frac{\partial\alpha_{ki}}{\partial S_j}+\frac{\partial\alpha_{ki}}{\partial\partial_x S_j}\partial_x\left(\frac{\alpha_{lj}}{\rho}\right)=0.
\end{eqnarray}
Equation~(\ref{eqSplit}) can be split into two terms with only one depending on $\rho$. To make the term that depends on $\rho$ vanish, we have to impose
\begin{equation}
\label{eq1}
\beta_{ij}\left[\partial_x\left(\frac{\partial\alpha_{lk}}{\partial\partial_x S_j}\right)-\frac{\partial\alpha_{lk}}{\partial S_j}\right]-\alpha_{kj}\frac{\partial\beta_{il}}{\partial S_j}+\frac{\partial\alpha_{il}}{\partial\partial_x S_j}\alpha_{kj}+\frac{\partial\alpha_{ki}}{\partial\partial_x S_j}\alpha_{lj}=0,
\end{equation}
for all $(i,l,k)$. In addition, canceling the term in Eq.~(\ref{eqSplit}) that  does not depend on $\rho$ leads to
\begin{equation}
\label{eq2}
\hspace*{-1.75 cm}\alpha_{ij}\left(\frac{\partial\alpha_{lk}}{\partial S_j}-\partial_x \frac{\partial\alpha_{lk}}{\partial\partial_x S_j} \right)+\alpha_{kj}\left(\frac{\partial\alpha_{il}}{\partial S_j}-\partial_x \frac{\partial\alpha_{il}}{\partial\partial_x S_j} \right)+\alpha_{lj}\left(\frac{\partial\alpha_{ki}}{\partial S_j}-\partial_x \frac{\partial\alpha_{ki}}{\partial\partial_x S_j} \right)=0,
\end{equation}
for all $(i,l,k)$. With these constraints, Eq.~(\ref{rnd7}) becomes
\begin{eqnarray}
\hspace*{-1.75 cm}\{F,\{G,H\}\}+&&\circlearrowleft_{(F,G,H)}=\int\Bigg\{\partial_x\left(\frac{F_i}{\rho}\right)\frac{1}{\rho}\partial_x\left(\frac{G_k}{\rho}\right)\frac{H_l}{\rho}\Bigg[\beta_{ij}\frac{\partial\beta_{kl}}{\partial S_j}\nonumber \\ 
&&\hspace{6 cm} -\beta_{ij}\frac{\partial\alpha_{kl}}{\partial\partial_x S_j}-\beta_{kj}\frac{\partial\alpha_{li}}{\partial\partial_x S_j}\Bigg]\nonumber \\
\label{eq66}
&&+\partial_x\left(\frac{F_i}{\rho}\right)\frac{1}{\rho}\partial_x\left(\frac{H_l}{\rho}\right)\frac{G_k}{\rho}\left[\beta_{lj}\frac{\partial\beta_{ik}}{\partial S_j}-\beta_{ij}\frac{\partial\alpha_{kl}}{\partial\partial_x S_j}-\beta_{lj}\frac{\partial\alpha_{ik}}{\partial\partial_x S_j}\right]\\
&&+\frac{1}{\rho}\partial_x\left(\frac{G_k}{\rho}\right)\partial_x\left(\frac{H_l}{\rho}\right)\frac{F_i}{\rho}\left[\beta_{kj}\frac{\partial\beta_{li}}{\partial S_j}-\beta_{lj}\frac{\partial\alpha_{ik}}{\partial\partial_x S_j}-\beta_{kj}\frac{\partial\alpha_{li}}{\partial\partial_x S_j}\right]\Bigg\}\ \mathrm{d}x.\nonumber
\end{eqnarray}
Choosing $F=\rho S_i$, $G=\int\rho S_k\ \mathrm{d}x$ and $H=\int\rho S_lx\ \mathrm{d}x$ leads to
$$
-\partial_x\left[\frac{1}{\rho}\left(\beta_{lj}\frac{\partial\beta_{ik}}{\partial S_j}-\beta_{ij}\frac{\partial\alpha_{kl}}{\partial\partial_x S_j}-\beta_{lj}\frac{\partial\alpha_{ik}}{\partial\partial_x S_j}\right)\right]=0,
$$
which has to be satisfied for any $\rho$, and therefore
\begin{equation}
\label{eq3}
\beta_{lj}\frac{\partial\beta_{ik}}{\partial S_j}-\beta_{ij}\frac{\partial\alpha_{kl}}{\partial\partial_x S_j}-\beta_{lj}\frac{\partial\alpha_{ik}}{\partial\partial_x S_j}=0,
\end{equation}
for all $(i,l,k)$. With this additional constraint, Eq.~(\ref{eq66}) is always satisfied, which proves that Eqs.~(\ref{eq1}), (\ref{eq2}),  and (\ref{eq3}) are necessary and sufficient conditions for Bracket~(\ref{brackFourFields}) to satisfy the Jacobi identity.

By introducing the expressions of $\alpha$ and $\beta$ given by Eqs.~(\ref{eq:fluidA}) and (\ref{eq:fluidB}) into  Eqs.~(\ref{eq1}), (\ref{eq2}),  and (\ref{eq3}), we end up with the following constraints~:
\begin{eqnarray}
\label{constraintsJacobi1}
&& \Gamma_{iklm}=\Gamma_{ilkm},\\
\label{constraintsJacobi2}
&& \Delta_{ikl}=\Delta_{lki},
\end{eqnarray}
where
\begin{eqnarray*}
\Gamma_{iklm}&=&\delta_m^k\left[(1-i-l)S_{i+l-2}+jS_{j-1}\frac{\partial S_{i+l-1}}{\partial S_j}\right]\\
&&-\delta_m^{k-1}\left[(i+l)S_{i+l-1}-(j+1)S_j\frac{\partial S_{i+l-1}}{\partial S_j}\right]-\frac{\partial S_{i+l-1}}{\partial S_j}\frac{\partial S_{k+j-1}}{\partial S_m},
\end{eqnarray*}
and
\begin{eqnarray*}
\Delta_{ikl}&=&\frac{\partial S_{i+k-1}}{\partial S_j}\left[(l+j)S_{l+j-1}-j(l+1)S_lS{j-1}-l(j+1)S_jS_{l-1}\right]\\
&&+l(i+k)S_{l-1}S_{i+k-1}+(l+1)(i+k-1)S_lS_{i+k-2}.
\end{eqnarray*}

\end{document}